\documentclass[twocolumn]{aa}
\usepackage{txfonts}
\usepackage{graphicx}
\def\note #1]{{\bf #1]}}
\def\dd{{\rm d}}
\def\xih{\xi_{\rm h}}

\begin{document}
\title{Probing the internal magnetic field of slowly pulsating B-stars through g modes }

\author{S.S. Hasan\inst{1} \and J.-P. Zahn\inst{2} \and J.
Christensen-Dalsgaard\inst{3}}
\offprints{S. S. Hasan, \email{hasan@iiap.res.in}}
\institute{Indian Institute of Astrophysics, Koramangala, Bangalore-560034, India
\and LUTH, Observatoire de Paris, F-92195 Meudon, France
\and Institut for Fysik og Astronomi, Aarhus Universitet, DK 8000 Aarhus C,
Denmark}

\date{Received 29 July 2005}

\abstract
{}
{We suggest that high-order g modes can be used as a
probe of the internal magnetic field of SPB (slowly pulsating B)
stars. The idea is based on earlier work by the authors
hich analytically investigated the
effect of a vertical magnetic field on p and g modes 
in a  plane-parallel isothermal stratified atmosphere.
It was found that even a
weak field can significantly shift the g-mode frequencies --- the
effect increases with mode order.} 
{In the present study we adopt the
classical perturbative approach to estimate the internal field of a 4
solar mass SPB star by looking at its effect on a low-degree ($l=1$)
and high-order ($n=20$) g mode with a period of about 1.5 d. }
{We find that a polar field strength of about 110 kG on the edge of the
convective core is required to produce a frequency shift of 1\%.
 Frequency splittings of that order have been observed in several SPB
variables, in some cases clearly too small to be ascribed to rotation.
We suggest that they may be due to a poloidal field
with a strength of order 100 kG, buried in the deep interior of the star.}
{}


\maketitle

\section{Introduction} It is well known that a magnetic field produces
a frequency splitting of stellar oscillations (Ledoux \& Simon 1957).
This effect has been extensively studied, using a perturbative
approach (e.g.  Goossens 1972; { 1976a,b; 1977; Goossens, Smeyers
\& Dennis 1976;} and more recently by  Gough \& Taylor 1984;
Dziembowski \& Goode 1984, 1985; Gough \& Thompson 1990; Shibahashi \&
Takata 1993).  
This method must be handled with care near the surface, where the
magnetic pressure dominates over the gas pressure, and where the
acoustic modes therefore strongly couple with the Alfv\'en modes (e.g.
Biront et al. 1982, Roberts \& Soward 1983, Campbell \& Papaloizou
1986, Dziembowski \& Goode 1996); such regions are better treated
using a non perturbative treatment (Bigot et al. 2000).

In an earlier paper Hasan and Christensen-Dalsgaard (1992) analytically
determined the frequency shift of p and g modes in an isothermal plasma due to
a homogeneous vertical magnetic field.  Using the full MHD equations, they
found that even a weak field (more precisely when $\beta \gg 1$, where $\beta$
is the ratio of gas to magnetic pressure) can produce a significant shift of
g-mode frequencies, while the effect on the p-mode spectrum is comparatively
small.  In principle this means that g-mode frequencies offer a diagnostic to
probe the internal field of stars in which g modes have been observed on the
stellar surface. 

Extensive observation campaigns have uncovered the existence of a
class of variable stars known as slowly pulsating B (SPB) stars which
are multiperiodic typically over a time scale of days (Waelkens 1991;
De Cat et al. 2005 and references therein).  These pulsations have
been identified with low degree $l$ (typically $l=1$ and 2) g modes of
high order, that are excited by the $\kappa$ mechanism in the metal
opacity bump at a temperature of about $2\times 10^5$~K (Dziembowski
et al. 1993).  These modes often occur in multiplets with closely
spaced periods (with a typical separation of 1\%). In some cases this
separation can clearly not be due to rotational splitting, which would yield
much larger spacings,  as was pointed out by De Cat and Aerts (2002).
In this letter we propose that such frequency splittings are due to the
presence of a magnetic field. If this hypothesis is correct, then the
{ splitting of frequencies} can be used to estimate the field strength in
the interior of SPB stars. 
 
\section{Magnetic frequency splitting of g modes}
As was established in the early papers quoted above, the frequency 
shifts  $\delta \omega$ due to a magnetic field are given by:
\begin{equation}
{\delta \omega \over \omega} = 
{1 \over 8 \pi \omega^2} {\int - \left[  (\nabla \times {\bf B'}) \times {\bf B}
+ (\nabla \times  {\bf B}) \times {\bf B'} \right] \cdot \xi^* \dd V \over
\int (\xi_r^2 + l (l+1) \xih^2) \rho  \dd V } \; ;
\end{equation}
here $\omega$ is the angular frequency,
$\vec \xi$ is the Lagrangian displacement with radial and horizontal 
components $\xi_r$ and $\xih$, respectively, and
${\bf B'}=\nabla\wedge({\vec \xi} \wedge {\bf B})$ is the perturbation 
in the equilibrium magnetic field $\bf B$. We shall deal here with fields whose
energy is sufficiently small compared to the gravitational energy, so that we
may neglect the structural changes caused by the Lorentz force.

{ Let us consider a poloidal axisymmetric field of the form:
\begin{equation} 
{\bf B} = B_0  \left[2 b(x) \cos \theta,- {1 \over x} \partial_x(x^2 b) \sin \theta,0\right] \; , 
\end{equation} 
where $x=r/R$ is the normalized radial coordinate, $R$ is the radius
of the star and $b(x) = {\cal O}(1)$, so that $B_0$ characterizes the field strength.  The displacement
$\vec \xi$ can be expressed as:
\begin{equation} 
{\vec \xi}= 
\left[\xi_r Y_l^m(\theta, \phi),\xih \partial_\theta Y_l^m,
                   \xih\ {i m \over \sin \theta}\, Y_l^m\right]  \exp i \omega t\; ,
\end{equation} 
where $Y_l^m$ is the spherical harmonic of degree $l$ and azimuthal
order $m$ (henceforth, denoted as $Y$).  
For a poloidal field given by equation (2), the perturbed field ${\bf
B'}$ is: 
\begin{eqnarray*}
B'_r & = & 
{B_0  \over R} \left[ \left[   l(l+1) \cos\theta\, Y + 
  \sin\theta\,\partial_\theta Y \right]  {2 \xih b \over x}   \right.\\
& & \left. - {1 \over \sin\theta}\,
\partial_\theta (\sin^2 \theta\,  Y ) \, {\xi_r \over x^2} \partial_x(x^2 b)  \right] \; , \\
B'_\theta & = & {B_0  \over R} 
 \left[ \cos\theta \, \partial_\theta Y \, {1 \over  x} \partial_x (2 x  \xih b) 
 - \sin\theta\, Y {1 \over  x} \partial_{x} \left( \xi_r  \partial_x (x^{2} b) \right)   \right.\\
 & & \left. \phantom{ {1 \over  x} \partial_x (x^{-2} \xi_r) \sin\theta Y} 
 - {m^2 Y\over \sin\theta} \, { \xih \over x^2} \partial_x(x^2 b)  \right]
\; ,\\
B'_\phi &=& im {B_0  \over R}  \left[ {\cos\theta \over \sin\theta}\,Y \, {1 \over x}  \partial_x (2 x  \xih b) 
-   \partial_\theta\, Y \, { \xih  \over  x^2} \partial_x (x^2 b)\right] \; .\\
\end{eqnarray*}
For g modes of high radial order $n$, it is straightforward to show that
in equation (1), the dominant term in the integrand of the numerator is the first one:
\begin{eqnarray}
\label{ratio}
& -& \left[(\nabla \times {\bf B'}) \times {\bf B}\right] \cdot \xi^* = \\
&& |{\bf B'}|^2 \simeq \left({B_0 \over R}\right)^2 \left|{2 \over x} 
{\dd \over \dd x} (x b \xih)\right|^2
\left[\left|\cos\theta {\partial Y \over \partial \theta}\right|^2 + m^2 \left|
{\cos\theta \over \sin\theta} Y \right|^2 \right] \; , \nonumber
\end{eqnarray}
whereas in the denominator it is $|\xih|^2$, since $\xih \gg \xi_r$
for $n \gg 1$.  In integrating by parts, we have neglected the surface
terms; this region anyhow requires a special, 
non-adiabatic and non-perturbative treatment
(see for instance Bigot et al.  2000).
From equation (1) it therefore follows that}
\begin{equation}
{\delta \omega \over \omega} = {1 \over 8 \pi \omega^2} {B_0^2 \over
\rho_{\rm c} R^2} C_{l,m}\, {\cal I} \; ,
\label{delsigma}
\end{equation}
where $\rho_{\rm c}$ is the central mass density,
\begin{equation}
{\cal I}\, ={\int \left| {2 \over x} {\dd \over \dd x}
(x b \xih)\right|^2 x^2 \dd x \over
\int  |\xih|^2 ( \rho/\rho_{\rm c}) x^2 \dd x } \; ,
\label{delsigma1}
\end{equation}
and 
\begin{equation}
C_{l,m} = 
{\displaystyle \int \left[\left|\cos\theta 
{\partial Y \over \partial \theta}\right|^2 
+ m^2 \left| {\cos\theta \over \sin\theta} Y \right|^2 \right] 
\sin\theta\, \dd \theta \over
l(l+1) \int |Y|^2 \sin\theta \, \dd \theta} \; .
\end{equation}
Obviously these constants do not depend on the sign of $m$: the
magnetic field reduces the $(2l+1)$ degeneracy of the eigenmodes to
only $l+1$, as already pointed out by Ledoux \& Simon (1957).  For $l=1$
and 2, we find:
\begin{equation}
C_{1,0} =  {1 \over 5} \; , \quad C_{1,1} = C_{1,-1} = {2 \over 5} \; ,
\end{equation}
\begin{equation}
C_{2,0} =  {9 \over 21} \; , \quad C_{2,1} =  C_{2,-1} = {8 \over 21} \; , 
\quad C_{2,2} =  C_{2,-2} = {5 \over 21} \; .
\end{equation}
\begin{table}
\caption{Frequency and periods for g modes of different radial orders
and degree $l=1$, for a 4 M$_\odot$ star with an age of 94 Myr.}
\label{table:1}
\centering
\begin{tabular}{c c c c c c c} 
   \hline\hline \\ 
  & \multicolumn{2}{c}{$l=1$}& &\multicolumn{2}{c}{$l=2$}\\
\cline{2-3} \cline{5-6} \\
$n$ &  $\nu$ ($\mu$Hz) & P (d) & $n$ & $\nu$ ($\mu$Hz) & P (d)  \\  \\ 
   \hline \\      
    23  &     6.683 & 1.732 & 40 & 6.761 &1.712   \\         
    20  &     7.728 & 1.498 & 35 & 7.730 &1.497  \\         
    15  &     9.983 & 1.159 & 30 & 9.035 &1.281  \\                  
    10  &     14.68 & 0.788 & 25 & 10.86 & 1.066 \\                 
    5   &     29.31 & 0.395 & 10 & 25.15 & 0.460  \\                 
       \hline \\ 
\end{tabular} 
\end{table}
It is convenient to express equation (\ref{delsigma}) as:
\begin{equation}
{\delta\omega \over \omega} = S_{\rm c}\, B_0^2, \; {\rm where} \, 
S_{\rm c} = {C_{lm}\ {\cal I}\over 8\pi\omega^2\rho_{\rm c} R^2} \; ,
\label{split}
\end{equation}
which we henceforth refer to as the splitting coefficient. This coefficient increases rapidly with period, since
${\cal I}$ increases also. 
{ Note that a toroidal field would produce a much lesser splitting,
since the leading term in equation (\ref{ratio}) would then be $|\xih|^2$
instead of $|\partial_x \xih|^2$.}

Finally, let us recall that the rotational splitting for g modes of high order is given by
\begin{equation}
{\delta\omega_{\rm rot}} = 
- m \left[1 - {1 \over l(l+1)} \right] \bar\Omega \; ,
\label{split-rot}
\end{equation}
where we have again assumed that $|\xi_r| \ll |\xih|$; 
we see that the frequency spacing
is of the order of the average angular velocity $\bar\Omega$.

\section{Model}
Following Dziembowski et al. (1993), we consider a model SPB star with
the following parameters: $M=4M_\odot$, $\log(L/L_\odot)=2.51$, $\log
T_{\rm eff}=4.142$, $X_{\rm c}=0.37$.  An equilibrium model for such a
star was calculated using the Aarhus Stellar Evolution Code (ASTEC)
(e.g. Christensen-Dalsgaard 1982, 1993); this used the Eggleton et al.
(1973) equation of state and OPAL opacities (Iglesias \& Rogers 1996)
and ignored diffusion and settling.  The evolved star had an age of 94
Myr.  We calculated the g modes of the above star using the Aarhus
adiabatic oscillation package (e.g. Christen\-sen-Dalsgaard \&
Berthomieu 1991).  Table 1 lists the cyclic frequencies $\nu$ and
periods $P$ for high-order  g modes corresponding to $l=1$ and $2$. 

\begin{figure}
\resizebox{\hsize}{!}{\includegraphics{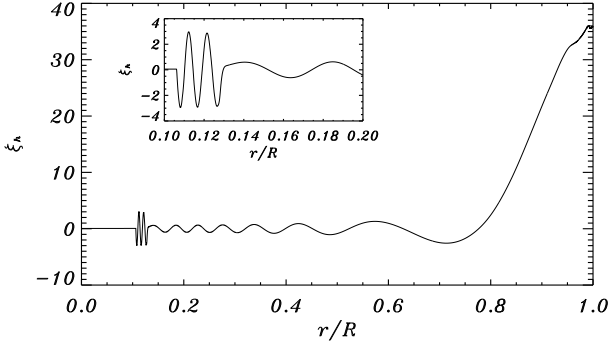}}
\caption{Horizontal component of the eigenfunction $\xih$  
(scaled such that $\xi_r = 1$ at the surface)
as a function of  normalized radius ($r/R$) for a $g_{20}^1$ mode 
with a period of 1.5 d for a 4 M$_\odot$ SPB star of 94 Myr. The inset
shows an enlargement of the region close to the edge of the convective core,
where the eigenfunction displays strong oscillations due to the presence of 
a steep composition gradient.}
\label{fig1}
\end{figure}

We first evaluate the frequency shift due to a magnetic field for the
$g_{20}^1$ mode (i.e. a g mode of radial order 20 and degree $l=1$),
which has a period of 1.5 d (a typical period for a SPB star). The
horizontal eigenfunction ($\xih$) of this mode is shown in Figure 1:
the radial component $\xi_r$ is normalized to unity on the surface
$r=R$ of the star.

{ We calculated the numerator of equation (\ref{delsigma1})  with different
functional forms $b(x)$ for the magnetic field. When the field is
constant throughout the star, the main contribution to that integral
comes from above $x \approx 0.8$, where $\xih$ has its largest
amplitude. But when the field is buried below that depth, or when it
tapers off as $b \propto x^{-q}$ with $q>1$, its main contribution
originates from a small region just above the convective core at $x_c
= 0.106$, where there is a steep gradient in the molecular weight in
this evolved star.  When we choose $b(x) = (x / x_c)^{-q}$, the
results depend little on $q$; those presented hereafter were obtained
with $q=3$.}

For this $g_{20}^1$ mode $\omega^2$ = $2.36\times 10^{-9}$ s$^{-2}$.  Using
$C_{1,1}-C_{1,0} = 1/5$, one finds $S_{\rm c} = 2.278 \times 10^{-6}$.  From
equation (\ref{split}), one deduces that in order to produce a 1\%  frequency
shift in $\delta \omega/\omega$, the polar field just above the convective core
has to be $B_{\rm pol} \simeq$ 110 kG. 

For comparison, we consider now the effect of the field on a g mode of order 10
and the same degree $l = 1$.
The horizontal component of the eigenfunction for this
mode is shown in Figure 2. In this case, $B_{\rm pol} \simeq$ 1100 kG, which is
an order or magnitude larger than for the $n=20$ mode with the same degree. 

Tables 2 and 3 give the splitting coefficient $S_{\rm c}$ for $l=1$
and $l=2$ g modes of various orders, and the corresponding polar
field strength $B_{\rm pol}$ required to produce a 1\% frequency shift.
In Table 3 we separate the contributions due to $C_{20}-C_{22}$ and
$C_{21}-C_{22}$ terms which produce different frequency shifts.  Note
how rapidly $S_{\rm c}$ increases with radial order $n$. 

\begin{table}
\caption{Splitting constant ($S_{\rm c}$) and value of the polar field
($B_{\rm pol}$), at the edge of the convective core for $l=1$ g modes that
would produce a frequency splitting of $1\%$, for a 4 M$_\odot$ star with
an age of 94 Myr.}
\label{table:2}
\centering
\begin{tabular}{c c c c} 
   \hline\hline \\ 
Mode & P (day) & $S_{\rm c}$ (Gauss$^{-2}$)& $B_{\rm pol}${\rm (kG)} \\ 
   \hline \\      
$g_{20}^1$ & 1.497 & $2.278\times 10^{-6}$ & 111 \\
$g_{15}^1$ & 1.159 & $3.765\times 10^{-7}$ & 274 \\
$g_{10}^1$ & 0.789 & $2.321\times 10^{-8}$ & 1102 \\
       \hline \\
\end{tabular} 
\end{table}
\begin{table}
\caption{Splitting coefficient ($S_{\rm c}$) and value of the polar field
($B_{\rm pol})$, at the edge of the convective core for $l=2$ g modes
corresponding to different transitions that would produce a frequency
splitting of $1\%$, for a 4 M$_\odot$ star with an age of 94 Myr.}
\label{table:3}
\centering
\begin{tabular}{c c c c c c c} 
   \hline\hline \\ 
& &\multicolumn{2}{c}{$C_{2,0}-C_{2, |2|}$}
                           &  &\multicolumn{2}{c}{$C_{2, |1|}-C_{2, |2|}$} \\
\cline{3-4} \cline{6-7} \\
 \hskip-6pt Mode &  P (day) & $S_{\rm c}$ (G$^{-2}$)& \hskip-12pt B$_{\rm pol}${\rm (kG)}  \hskip-6pt  & &  \hskip-6pt $S_{\rm c}$ (G$^{-2}$)& \hskip-18pt  B$_{\rm pol}${\rm (kG)}  \\
   \hline \\      
$g_{35}^2$ & 1.497 & $6.435\times 10^{-7}$ &210  & & $4.827\times 10^{-7}$  & 242   \\
$g_{30}^2$ & 1.281 & $3.422\times 10^{-7}$ &288  & & $2.566\times 10^{-7}$  & 332   \\
$g_{10}^2$ & 1.066 & $7.897\times 10^{-9}$ & 1890  & & $5.922\times 10^{-9}$  &2180   \\
       \hline \\
\end{tabular} 
\end{table}

\begin{figure}
\resizebox{\hsize}{!}{\includegraphics{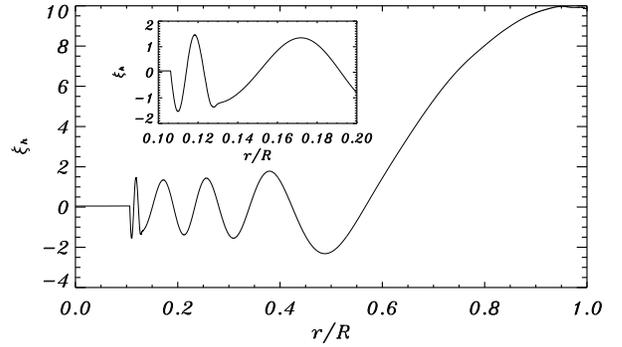}}
\caption{Same as in Fig. 1, but for the $g_{10}^1$ mode. The inset
shows an enlargement of the region close to the convective core.}
\label{fig2}
\end{figure}

\section{Discussion} Our calculations suggest that high-order g modes
can provide a sensitive diagnostic of the internal magnetic field in
SPB stars, at least when these are sufficiently evolved. The main
contribution to the splitting coefficient then comes from a small
region just above the convective core (close to $x=0.106$ in a 4
$M_\odot$ star of 94 Myr), where a steep helium gradient has built up
due to the shrinking of the convective core.  This causes a sharp peak
in the Brunt-V\"ais\"al\"a frequency $N$, which is reflected in a
series of closely spaced nodes in the horizontal eigenfunction.  In
order to examine the impact of such a region, we repeated the
calculation for a ZAMS star with the same chemical composition and
physical parameters as our evolved SPB star, and found indeed that the
frequency splitting was then much less sensitive to the strength of
the magnetic field.

We concentrated on g modes of low degree and high order, which are
typically excited in SPB stars. We found that a polar field of 110 kG
in the vicinity of the convective core causes a splitting of 1\% for a
g$^1_{20}$ mode.  For a simple dipolar configuration of the magnetic
field, this would translate into a 120 G polar field on the surface of
the star; { we quote this figure only for illustration purpose,
since there is no way to deduce the surface field from the deep field.
We have checked that the rest of the star contributes little to the
splitting, provided that the field is buried below the depth of about
$x=0.80$, or that it tapers off at a faster rate than $b \propto 1/x$.
In such a situation we find that high-order g-modes can be be used to
probe the deep interior field. This result  is not sensitive to the
precise field configuration.}  

With moderate- to low-order modes, the diagnostic is much less
sensitive.  For a $l=2$, $n=10$ g mode with a period of 1.1~d, a polar
field of about 2 MG would is required to produce a frequency splitting
of 1\%.

So far only one SPB star has been detected with a magnetic field.  Recently,
Neiner et al.  (2003) have reported the discovery of a field on the SPB star
$\zeta$ Cas, in which a non-radial pulsation with a period $P=1.56$~d was
detected. A field strength for the time-averaged line of sight polar component
of $330^{+120}_{-65}$~G was inferred. If this field is the visible part of a
deeply rooted magnetic field, it could also leave a signature in the splitting
of high-order g modes.  {We should emphasize that according to our results,
a field of order 100 kG at the edge of the convective core would be required to
produce  a 1 \% splitting in typical g modes.}

We considered here only a purely poloidal configuration, { similar
to many earlier papers quoted in Sect. 1.} Such a configuration is
known to be unstable. As was shown by Tayler and collaborators (Tayler
1973, Pitts \& Tayler 1985), and as was illustrated recently by the
numerical simulations of Braithwaite and Spruit (2004), the
configurations which are likely to resist non-axisymmetric MHD
instabilities are combinations of large-scale toroidal and poloidal
fields of about equal strength. Taking this into account is unlikely
to alter our conclusions concerning the detectability of the deep
magnetic field, because the frequency splitting would be much more
sensitive { to the poloidal than to the toroidal component in such
a combined field.}

\begin{acknowledgements}
{ We thank the referee for comments clarifying the role of the global
properties of the magnetic field.}
S.S. Hasan and J.-P. Zahn are grateful to the Indo-French Centre
for the Promotion of Advanced Research, New Delhi for supporting this
project through grant number 2504-3.  We thank M. J. Thompson for
useful discussions.  J. Christensen-Dalsgaard acknowledges the
hospitality of the Indian Institute of Astrophysics, Bangalore, and
the High Altitude Observatory, Boulder, CO, U.S.A. during this project. 
\end{acknowledgements}


\end{document}